\documentclass[preprint,prb]{revtex4}

\usepackage{verbatim}
\newcommand{\lb}{\langle}
\newcommand{\rb}{\rangle}
\newcommand{\half}{\frac12}
\usepackage{graphicx}
\usepackage{amssymb}
\begin{document}

\title{Understanding the temperature and the chemical potential
using computer simulations}

\author{Jan Tobochnik}
\email{jant@kzoo.edu}
\affiliation{Department of
Physics, Kalamazoo College, Kalamazoo, Michigan 49006}

\author{Harvey Gould}
\email{hgould@clarku.edu}

\affiliation{Department of Physics, Clark University, Worcester,
Massachusetts 01610}

\author{Jonathan Machta}
\email{machta@physics.umass.edu}

\affiliation{Department of Physics, University of
Massachusetts, Amherst, Massachusetts 01003\\
Received Date: 10/13/04  Accepted Date: 10/26/04  Index Code: 2.70, 5.20,
5.70}

\begin{abstract}
Several Monte Carlo algorithms and applications that are useful
for understanding the concepts of temperature and chemical potential are
discussed. We then introduce a generalization of the demon algorithm that
measures the chemical potential and is suitable for simulating systems with
variable particle number.
\end{abstract}

\maketitle

\section{INTRODUCTION}

There are several physical quantities in thermal physics that are
difficult to understand because their operational definitions
involve sophisticated reasoning. For example, what is temperature
and what is the best way to introduce it? One way to introduce the
temperature is to invoke kinetic theory or the equipartition
theorem and take the temperature to be proportional to the mean
kinetic energy per particle of a system. However, the use of simple kinetic
theory involves many approximations, which can lead to doubt and
confusion in the minds of many students as to the generality of the
relation between the temperature and kinetic energy. Others will
say that temperature is a measure of internal energy, and some will
define the temperature as what a thermometer measures. However, the
latter definition requires a long and subtle argument
about how the change in some property of a thermometer (such as the
volume of a column of mercury) can be calibrated.

To understand the role of temperature, it is necessary to realize
that when two objects at different temperatures are brought into
thermal contact, energy is transferred from the hotter object to
the colder object until the temperature of the two objects becomes
the same. Temperature is a measure of how easy it is for an object
to give up its energy. The higher the temperature, the easier it is. It
is a fact of nature that this single quantity determines
the direction of energy transfer in systems near equilibrium. 

The nature of the chemical potential is even more obscure for beginning 
students of statistical physics.\cite{baierlein} Unlike temperature, the
chemical potential is not familiar from everyday life. The chemical
potential determines the direction of transfer of particles between
systems that can exchange particles. The higher the chemical potential,
the easier it for an object to give up particles.

When discussing abstract concepts, it is useful to think in terms
of a simple model in order to strip away complicating factors that
accompany actual measurements but are tangential to conceptual
understanding. For example, we use ideal clocks and meter sticks
when discussing time and space. For this reason, it would be useful
to have an ideal thermometer.

In this paper we will introduce an ideal thermometer in the context
of computer simulations of thermal systems. We also will discuss a
new simulation algorithm that can be used to measure the chemical
potential. The latter is based on a generalization of the demon
algorithm introduced by Cruetz.\cite{creutz}

\section{The Demon Algorithm}

Computer simulations provide a vehicle for understanding many of the key
concepts in statistical mechanics. Most simulations are based on either Monte
Carlo or molecular dynamics algorithms.\cite{gt} In Monte Carlo simulations
the configurations are sampled according to the appropriate probability
distribution,
and averages are computed from these configurations. One of the
most influential Monte Carlo algorithms is the Metropolis
algorithm whose 50th anniversary was celebrated in 2003.\cite{los
alamos} This algorithm simulates a system at constant temperature
$T$, volume $V$, and number of particles $N$, that is, it samples
configurations in the canonical ensemble. We summarize it here so
that the reader can see how it relates to the demon algorithm which
we will describe next.

\begin{enumerate}

\item Begin with an arbitrary initial configuration with the desired
values of $V$ and
$N$. If the system consists of particles, the configuration would
specify the positions of the particles. If the system were a lattice of
spins, the configuration would consist of the orientation of each spin.

\item Make a trial change such as moving a particle
at random or flipping a spin and compute the change in energy,
$\Delta E$.

\item If $\Delta E \leq 0$ or if $r \leq e^{-\beta \Delta E}$, where
$r$ is a random number uniformly distributed in the unit interval,
then accept the change. Otherwise, reject the change, but include the
unchanged configuration in the system averages. Here $\beta = 1/kT$ and $k$ is
Boltzmann's constant.

\item Repeat steps 2 and 3 and compute various averages and their
statistical uncertainties periodically. Allow the system to
equilibrate before computing averages.

\end{enumerate}
Once the system has reached equilibrium, the Metropolis algorithm
produces configurations that occur with a probability proportional
to the Boltzmann distribution.

In the Metropolis algorithm the temperature is an input parameter,
and we cannot obtain much insight into its nature by using this
algorithm. We now describe the demon algorithm for which the energy
is an input parameter and temperature is measured, thus providing a
way of thinking about the temperature. In this algorithm one extra
degree of freedom, called the demon, can exchange energy with the
system of interest, and the energy of the system plus the demon is
kept constant. The demon satisfies only one constraint, namely, its
energy, $E_d$, satisfies $E_d \geq 0$. Thus, the demon algorithm samples
configurations in the microcanonical or constant energy ensemble.

The steps of the demon algorithm can be summarized as follows:

\begin{enumerate}

\item Begin with an arbitrary initial configuration with the
specified parameters, $E$, $V$, and $N$. Obtaining such a
configuration is usually more difficult than determining an
initial configuration for the Metropolis algorithm. For simplicity,
we usually set
$E_d = 0$, but we may give it any (positive) value
such that the total energy $E$ has the desired value.\cite{comment}

\item Make a trial change in the system and compute the change in
energy, $\Delta E$.

\item If $\Delta E < E_d$, accept the change, and subtract $\Delta E$
from $E_d$. Otherwise, reject the change, but include the unchanged
configuration as the new configuration.

\item Repeat steps 2 and 3 as in the Metropolis algorithm. In
addition, accumulate data for the histogram of demon energies, $H(E_d)$,
after each Monte Carlo step per particle or lattice site.

\end{enumerate}

Note that an immediate advantage of the demon algorithm is that fewer
random numbers and no computationally expensive calculations of the
exponential function are needed. Thus, for some calculations the demon
algorithm can be much faster than the Metropolis algorithm.\cite{creutz}

We now discuss why the demon acts as an ideal thermometer. The latter
should be as small as possible so that it does not affect the system
of interest, and should have a convenient macroscopic property that
changes in a well defined way with the temperature. The demon
satisfies both conditions. If the system of interest has many
degrees of freedom, the demon is a small perturbation of order $1/N$ on the
system. As $N$ increases, the demon becomes less
significant, and thus the demon becomes an ideal measuring device. The
demon exchanges energy with the system, but does not change its energy
very much. 

To relate the properties of the demon to the temperature, we can go back to
the fundamental definition of temperature,
\begin{equation}
\label{eq:defT}
\frac{1}{T} = \frac{\partial S}{\partial E},
\end{equation}
where $S(E)$ is the entropy and $E$ is the energy of the system. The
entropy in the microcanonical ensemble is defined in terms of the number of
states of the system with energy $E$,
$\Omega(E)$:
\begin{equation}
S(E) = k \ln \Omega(E).
\end{equation}
The basic principle of statistical mechanics is that all states with
the same energy are equally likely, and because the demon can have only
one state for each of its possible energy values, it follows that the
probability of finding the demon with energy
$E_d$ is proportional to the number of states of the system at energy $E$,
\begin{equation}
P(E_d) \propto \Omega(E).
\end{equation}
Note that $E_d$ and $E$ are linearly related because $E+E_d$ is held fixed
during the simulation.

We can now determine the number of accessible states
in terms of the entropy and its derivative:
\begin{equation}
\Omega(E) = \exp[S(E)/k] = \exp\Big[S(E+E_d)/k - E_d \frac{\partial
S(E)}{\partial E} + \cdots \Big].
\end{equation}
In the limit $E_d \ll E$ we can ignore higher terms in the Taylor
expansion of the entropy. From the definition of the temperature,
Eq.~(\ref{eq:defT}), and the fact that $S(E+E_d)$ is a constant, we find,
\begin{equation}
\label{eq:ped}
P(E_d) \propto e^{-\beta E_d}.
\end{equation}
Thus a plot of
$\ln P(E_d)$ versus $E_d$ will be a straight line with a slope of
$-\beta$. Hence, the demon measures the temperature without disturbing
the system of interest. Moreover, the mean value
$\lb E_d \rb = kT$ for any dimension if the demon energy distribution
is continuous.\cite{footexplain} Note that Eq.~(\ref{eq:ped}) is simply the 
Boltzmann probability for a system at temperature $T$ to be in a
microstate with energy
$E_d$.

By thinking about how the demon works, we can develop an intuition
about temperature. For example, high temperatures or small $\beta$
implies that it is easy for the demon to obtain energy from the system,
and hence a plot of $\ln P(E_d)$ versus $E_d$ would have a shallow slope. At
low temperatures the system has difficulty giving up energy and thus the
slope of $\ln P(E_d)$ would be steep. We also can see that the
temperature is an intensive quantity. If we increase the size of the
system (doubling the volume and number of particles for example), the
internal energy would double, but the temperature would be unchanged.
From the perspective of the demon, the changes in the energy during
trial moves do not depend on the size of the system, and thus there is
no reason why the energy distribution of the demon should change, and
hence the temperature measured by the demon is independent of the
size of the system.

For many purposes the equipartition theorem provides a way of defining
or determining the temperature. The theorem states that for a
classical system in equilibrium with a heat bath at temperature $T$,
the mean value of each contribution to the total energy that is
quadratic in a coordinate in phase space is
$\half kT$. Thus, for an ideal monatomic gas of $N$ particles in three
dimensions, the total internal energy is given by $3NkT/2$. However,
students frequently do not appreciate that this relation holds only
for quadratic degrees of freedom and only for classical systems. The
equipartition theorem also is independent of the interactions, if any,
between the particles, a fact that is obscured by the usual derivations of
the theorem for an ideal gas.

\begin{figure}[h]
\begin{center}
\includegraphics[scale=0.5]{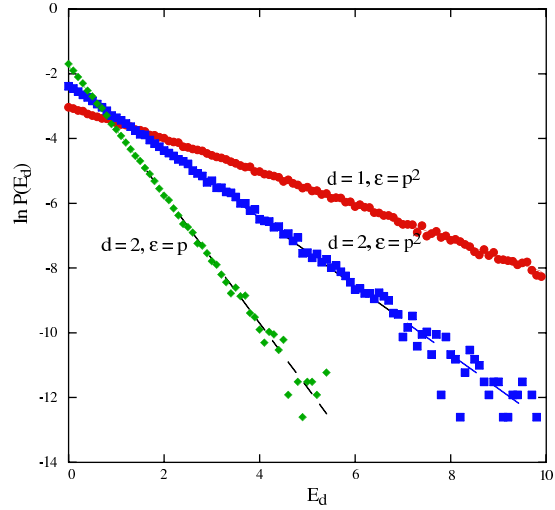}
\caption{\label{fig1}Plot of the demon energy distribution,
$P(E_d)$, for a demon in equilibrium with a classical ideal gas of
$N = 100$ particles in $d$ spatial dimensions. The trial moves are momentum
changes. Only 3000 Monte Carlo steps per particle were used. For $d=1$ and
the usual quadratic relation between particle energy and momentum, we
find $T = 1.93$ from the inverse slope of $P(E_d)$ and $\lb E
\rb/N =
0.980$, approximately a factor of two difference as would be expected
from the equipartition theorem. Similarly, for
$d=2$, we find
$T = 0.948$ and $\lb E \rb/N = 0.990$, which also is consistent with the
equipartition theorem. However, for $d=2$ and a linear dispersion
relation, we find $T =0.503$ and $\lb E \rb/N = 0.995$, which, as
expected, is not consistent with the equipartition theorem.}
\end{center}
\end{figure}

We can easily demonstrate the nature of the
equipartition theorem by applying the demon algorithm to the ideal
gas for which the kinetic energy of a
particle is given by
$\epsilon(p) = p^2/2m$. In this case the trial moves are changes in the momentum $p$ of a
particle. Figure~\ref{fig1} shows the demon probability distribution for an
ideal gas of $N=100$ particles in one and two dimensions. We find $\lb
E/N\rb = (1/2)kT$ for $d=1$ and
$\lb E/N\rb = kT$ for $d=2$, consistent with the predictions of the
equipartition theorem to within a few percent. The agreement could be
improved by running for more than only 3000\, Monte Carlo steps per
particle and by eliminating some of the higher energy results which have
more scatter. Also for these higher energies the demon energy
has become a significant fraction of the total energy, and hence the
Boltzmann distribution becomes less applicable because of finite size
effects. 
 
We can easily modify the simulation by assuming relativistic particles
so that the kinetic energy of a particle is proportional to its momentum,
that is,
$\epsilon \propto p$. In this case the demon algorithm still
produces a linear plot for
$\ln{P(E_d)}$ versus $E_d$, but the temperature derived from this
plot is not the same as that given by the kinetic energy per particle of
the system. Instead if $\epsilon \propto p$, each momentum degree of freedom
contributes $kT$ to the mean energy. For a
two-dimensional relativistic gas we find that $T=0.503$ from the demon
energy distribution (see Fig.~\ref{fig1}), but $\lb E\rb/N
= 0.995$, which is consistent with our expectation, but not with the usual
statement of the equipartition theorem.

Even though the internal energy is a function of temperature, these two
concepts play different roles in thermodynamics. Often students say that the
temperature is a measure of the internal energy. Some textbooks use the term
thermal energy to refer to
$kT$. However, it is less confusing to students if the difference between
the concepts of energy and temperature is emphasized. Temperature is a
measure of the ability of a system to take on energy by heating or cooling.
Clearly, it has a conceptual role that is distinct from any form of energy. 

\begin{figure}[h]
\begin{center}
\includegraphics[scale=0.5]{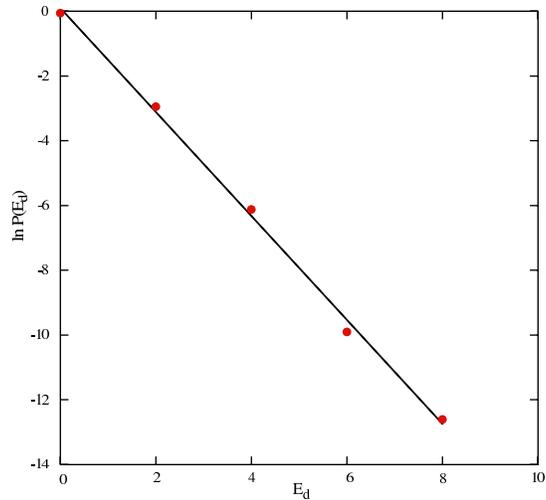}
\caption{\label{fig2} The demon energy
distribution, $P(E_d)$, for a one-dimensional Ising model with
$N=100$ and 3000\,mcs. The slope is $-1.60$ which implies that
$T=0.625$. The energy per particle of the Ising model is found to
be $\lb E\rb/N = -0.801$.} 
\end{center}
\end{figure}

There are many systems that have no kinetic energy at all and yet these
systems have a well defined temperature. For example, the Ising model of
magnetism assumes that the internal energy of a configuration of spins in
the absence of an external magnetic field is
\begin{equation}
E = - J\!\sum_{<ij>}s_is_j, \label{ising}
\end{equation}
where $J$ is the coupling constant, which is usually set equal to unity in a
simulation, and
$s_i = \pm 1$; the index $i$ denotes the lattice site. The notation
${<}ij{>}$ indicates that the sum is over nearest neighbor pairs of spins.
The demon algorithm can be applied to the Ising model, and indeed that was
its first use.\cite{creutz} Trial changes are made by choosing a spin at
random and attempting to flip it. In Fig.~\ref{fig2} we show a plot of
$\ln P(E_d)$ versus $E_d$ for the one-dimensional Ising model. Note that the
possible values of the demon energy are multiples of 2, because flipping a
spin can only change the energy by 0, 2, or 4.\cite{foot2} The value of the
slope of the fitted line in Fig.~\ref{fig2} implies that the temperature is
approximately 0.625. The mean system energy per spin is found to be
$-0.801$ (in units for which $J=1$). As is well known the energy increases
with temperature, but the dependence is not linear.

\begin{figure}[h]
\begin{center}
\includegraphics[scale=0.5]{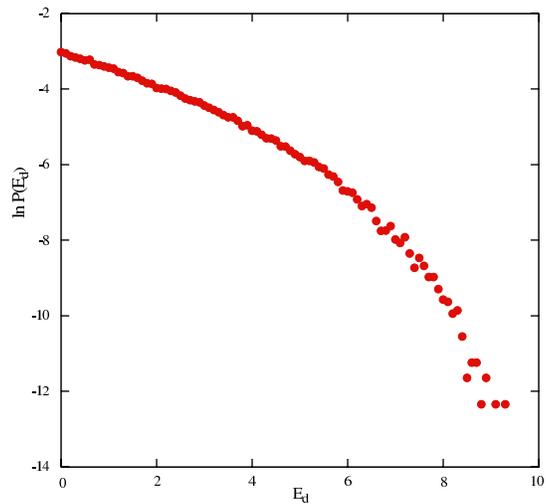}
\caption{Results for the demon energy distribution
for an one-dimensional ideal gas with $N=10$ particles showing the
strong curvature due to the demon having a significant fraction of
the total energy. Over 20,000\,mcs were used.} 
\label{fig3}
\end{center}
\end{figure}

As we have mentioned, the probability distribution of the demon is not a
perfect exponential for larger values of $E_d$ because of finite size
effects that exist in any  simulation. The reason goes back to the 
derivation of the Boltzmann distribution. In such a derivation the
system of interest is assumed to be  small so that its energy
fluctuations  are much smaller than the fluctuations of the energy of the heat bath
with which it is in thermal contact. In our simulations
the demon plays the role of the system of interest, and the simulated system
plays the role of the heat bath. As we have seen it is possible for the demon
to obtain a significant fraction of the energy of the system. An example of
this curvature can be seen in the plot in Fig.~\ref{fig3} of
$\ln{P(E_d)}$ for larger $E_d$; in this simulation
$N = 10$ and the mean energy of the demon is 1.68, which is 16.8\% of the
total energy of the system. 

Another desirable feature of the demon algorithm is that it can be used in
conjunction with other simulations. We could add the demon algorithm to a
Metropolis or MD simulation and measure the temperature with the demon and
compare it to other measures of the temperature.

\section{The chemical potential demon}

Just as temperature measures the ability of a system to transfer energy to
another system and pressure is a measure of the ability of a system to
transfer volume, the chemical potential is a measure of the ability of a
system to transfer particles. Two systems in thermal and particle
equilibrium will come to the same temperature and the same chemical
potential.

The chemical potential can be measured using the Metropolis algorithm
and the Widom insertion method.\cite{frenkel} From thermodynamics we
know that
\begin{equation}
\mu = \Big(\frac{\partial F}{\partial N}\Big)_{V,T} = -kT \ln
\frac{Z_{N+1}}{Z_N}
\label{dfdn}
\end{equation}
in the limit $N \to \infty$, where $F$ is the Helmholtz free energy
and
$Z_N$ is the
$N$-particle partition function. The ratio $Z_{N+1}/Z_N$ is the
average of
$e^{-\beta \Delta E}$ over all possible states of the added
particle with added energy
$\Delta E$. In a Monte Carlo simulation with Widom insertion we would compute the
change in the energy $\Delta E$ that would occur if an imaginary
particle were added to the $N$ particle system at random for many
configurations generated by the Metropolis algorithm. The chemical
potential is then given by
\begin{equation}
\mu = -kT
\ln \lb e^{-\beta \Delta E}\rb, \label{insertion}
\end{equation}
where the average $\lb \ldots \rb$ is over many configurations
distributed according to the Boltzmann distribution. Note that in the
Widom insertion method, no particle is actually added to the system 
during the simulation.

If we consider an ideal classical gas and consider only the momentum
degrees of freedom, then Eq.~(\ref{insertion}) would lead to the usual
analytical result for $\mu$ (see Eq.~(\ref{analytic})) if we include a
factor of 
$1/N$ in the argument of the logarithm to account for the
indistinguishability factor
$1/N!$ in
$Z_{N}$.
Because the contribution to the chemical potential due to the momentum
degrees of freedom is known exactly, only the
position degrees of freedom are retained in typical Monte Carlo
simulations of interacting systems, and the chemical potential in Eq.~(\ref{insertion}) is
interpreted as the excess chemical potential beyond the ideal gas
contribution.\cite{kofke}

Although the Widom insertion method gives some insight into the nature of
the chemical potential, it is not easy to obtain an intuitive understanding
of the chemical potential.\cite{baierlein}
For example, why is the chemical potential negative for a classical
ideal gas? A theoretical argument for the sign of the chemical potential
is the following.\cite{goodstein} If we add a particle to a closed system
at constant volume and energy, we have from the fundamental
thermodynamic relation
\begin{equation}
\Delta E = T \Delta S + \mu \Delta N = 0.
\label{eq:fund}
\end{equation}
{}From Eq.~(\ref{eq:fund}) we see that adding a particle with no change in
energy, $\Delta E = 0$ and $\Delta N = 1$, can be done only if $\mu = -
T \Delta S$. Because adding a particle at constant energy and volume can
only increase the entropy, we must have
$\mu < 0$. This argument breaks down for fermions at low temperatures because
the low energy states are already filled and thus $\Delta E \ne 0$.

To understand the role of the chemical potential, we have extended
the demon algorithm so that the demon also can hold particles. If we use
the thermodynamic definition of $\mu$,
\begin{equation}
\label{eq:defmu}
\frac{\mu}{T} = -\frac{\partial S(E,N)}{\partial N},
\end{equation}
and follow the derivation of Eq.~(\ref{eq:ped}),
we can show that the probability that the demon has energy
$E_d$ and
$N_d$ particles is given by:
\begin{equation}
\label{eq:macroprob}
P(E_d,N_d)
\propto e^{-\beta (E_d - \mu N_d)}. \label{gibbs}
\end{equation}
Thus, the slope of $\ln P(E_d, N_d)$ versus $N_d$ for fixed $E_d$ yields
$\beta \mu$ and the demon serves as an ideal measuring device for
the chemical potential. Note that Eq.~(\ref{eq:macroprob}) more generally
is the probability of a system at temperature $T$ and chemical
potential $\mu$ to be in a microstate with
$N_d$ particles and energy $E_d$.

For simplicity, we introduce a lattice in phase space, including the
momentum degrees of freedom, to do our simulations. The generalized demon
algorithm for such a lattice is as follows:

\begin{enumerate}

\item Begin with an arbitrary initial configuration with the desired energy
$E$ and number of particles $N$. One way to do so is to add one particle at
a time to random lattice sites, computing the total energy at
each step until we have added $N$ particles and reached the desired energy.
In the simple systems we will consider here, there are many zero energy
states so that we can simultaneously obtain the desired $E$ and $N$. Set
$E_d = 0$ and
$N_d = 0$.

\item Choose a lattice site in the system at random.

\item If there is a particle present at this site,
compute the change in energy $\Delta E$ that would result if the
particle were removed. If the demon has sufficient energy, then accept the
change, subtract $\Delta E$ from the demon, and increase $N_d$ by 1,
where $N_d$ is the number of particles held by the demon. Otherwise,
reject the move, but include the unchanged configuration as the new
configuration. Go to step 5.

\item If there is no particle present and the demon contains at least
one particle, attempt to place a particle at the chosen site by
computing the change in energy, $\Delta E$, needed to add a particle. If
$\Delta E < E_d$, then accept the move, subtract $\Delta E$ from
$E_d$, and let $N_d \to N_d - 1$. Otherwise reject the move, but include
the unchanged configuration as the new configuration.

\item Repeat steps 2--4 and compute various averages and their
statistical uncertainties as before. Accumulate data for $P(E_d, N_d)$
after each Monte Carlo step per lattice site.
\end{enumerate}

As can be seem, these modifications of the usual demon algorithm are
straightforward. For dilute systems there is no need to include particle
moves as in the original demon algorithm. These moves are not necessary
because every such move is equivalent to a particle removal at one site and
insertion at another. Thus, all possible configurations can be reached by
only using particle insertions and removals. However, for dense systems the
probability of adding a particle is so low that particle moves would become
necessary to obtain a reasonable exploration of phase space.

We can plot $\ln P(E_d, N_d)$ versus $E_d$ for a particular value of $N_d$
or versus
$N_d$ for a given value of $E_d$. The slope of the first plot is $-\beta$ and
the slope of the second is $\beta \mu$. Sometimes, there is some curvature
in the plot of $\ln P(E_d, N_d)$ versus $N_d$, which makes it difficult to
extract the chemical potential. This curvature occurs when the number of
particles in the demon becomes a significant fraction of $N$. 

\section{Simulation Results}\label{sec:results}

\subsection{Ideal Gas on a Phase Space Lattice}\label{sec:idealgas}

We first consider an ideal classical gas for which the positions and momenta
of the particles take on discrete values. For simplicity, we consider
one spatial dimension so that we can visualize the phase space of the
system as a two-dimensional lattice. The length of the lattice along the
position axis in phase space is the size of the system, $L$, but the
length in the momentum direction could be infinite. However, we need
only to make it large enough so that the probability of a particle
having a momentum larger than the extent of our momentum axis is
negligible. For our simulations we chose $L=1000$ and the momentum
axis to extend from $-10$ to 10. Visual inspection of the particles in
phase space showed that the particles were never near the ends of
the momentum axis, and thus the extent of the momentum axis was
sufficiently large. In order that all energies ($= p^2/2m$) be integers,
we set the mass
$m = 1/2$. In this way each energy index of the array for
$P(E_d,N_d)$ equals an unique value of the energy.

For simplicity, we impose the restriction that no two particles can
occupy the same lattice site in phase space. In the dilute limit, this
restriction will have a negligible effect on the results, but we
will show how the restriction changes our results for
high density, low energy systems. In analytical calculations phase space is
divided into cells, which is necessary to allow for the counting of states.
This imposition is further justified using arguments about the uncertainty
principle. A trial move consists of choosing a phase space cell or site at
random and either attempting to add a particle if it is empty or attempting
to give the particle to the demon if the site is occupied. Because we choose
a state not a particle, our algorithm automatically implies the
indistinguishability of the particles.

Our lattice model corresponds to a semi-classical model of a ideal gas
for which a factor of
$1/N!$ has been included to account for particle
indistinguishability.\cite{schroeder} This factor is only accurate in
the dilute limit because it allows more than one particle per state. 
 
As far as we know, this lattice model of an ideal gas has not been
simulated in an educational context, probably because most calculations can
be done analytically. However, by thinking about how the simulations work,
we believe that students can obtain a more concrete understanding. For
example, setting up phase space as a two-dimensional lattice and showing the
configurations can provide an animated visual representation which can help
students better understand the meaning of the analytical calculations.

\begin{figure}[h]
\begin{center}
\includegraphics[scale=0.5]{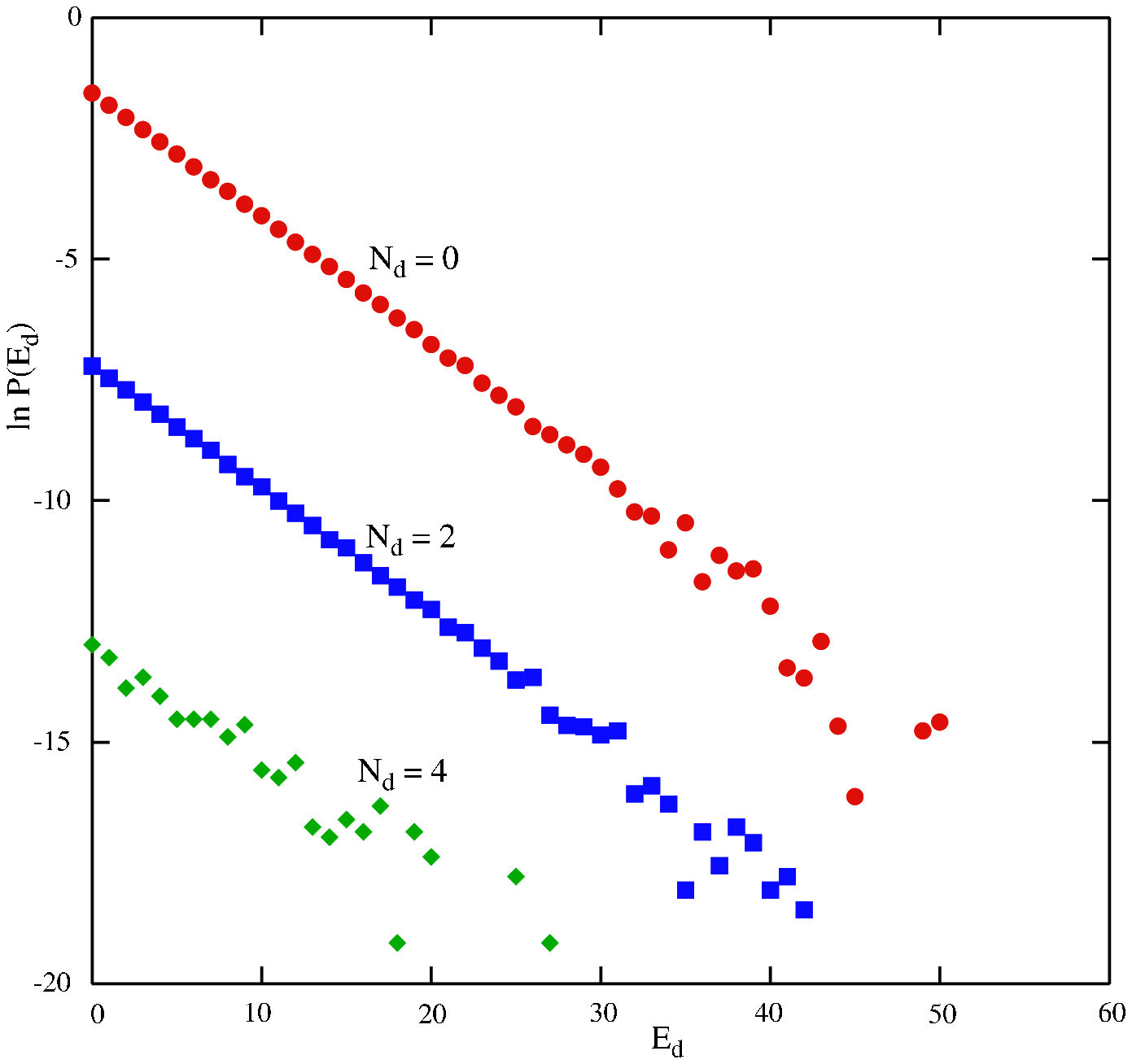}
\includegraphics[scale=0.5]{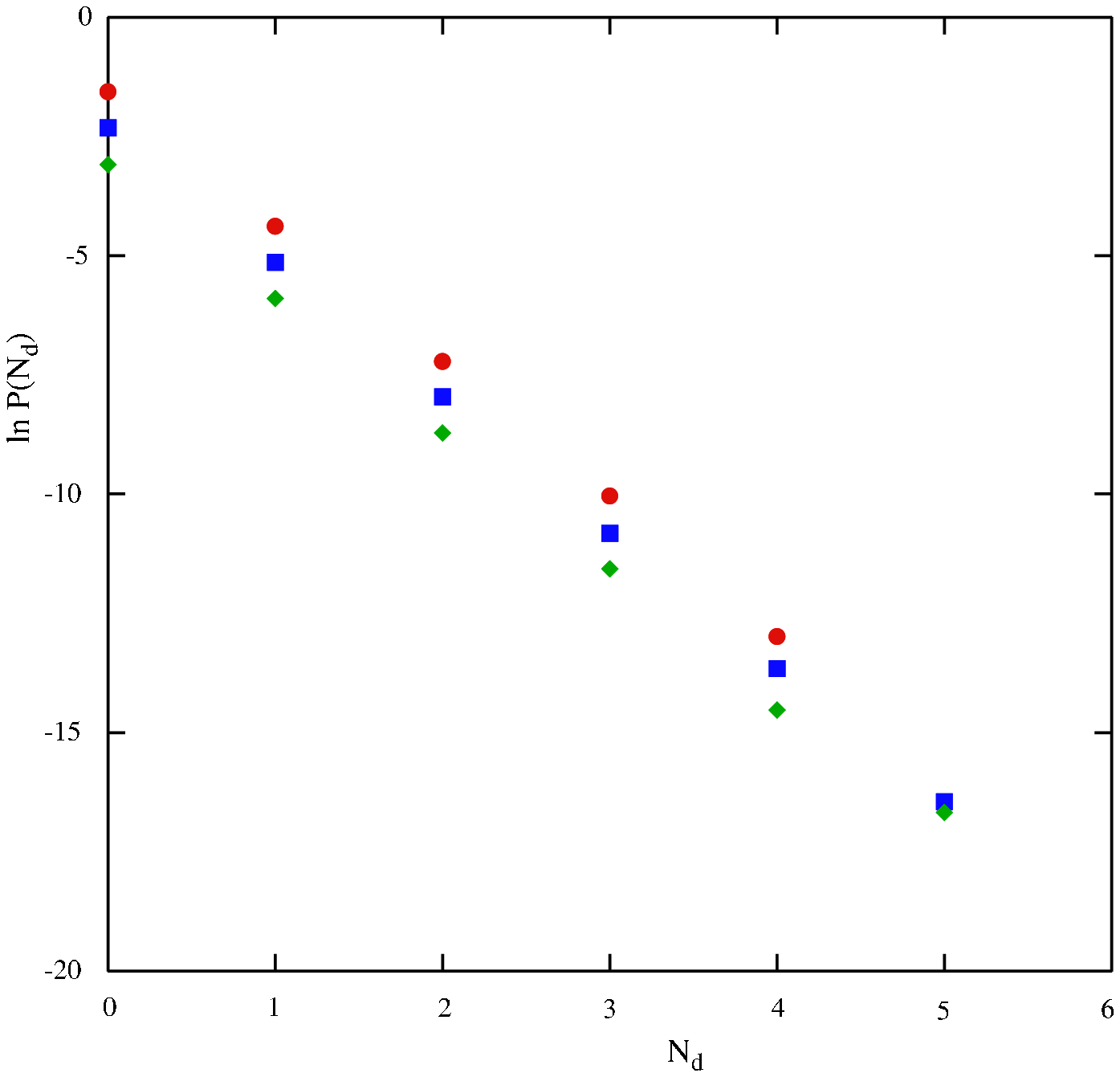}
\caption{\label{fig4} Results for a one-dimensional ideal classical gas on
a two-dimensional phase space lattice with $N=200$ and total energy
$E=400$. The system was equilibrated for 500\,mcs and averages were
taken over 10,000\,mcs. (a) The demon energy distribution,
$P(E_d,N_d)$, for
$N_d = 0$, 2, and 4 with slopes of $-0.26$, $-0.25$, and $-0.25$
respectively. (b) The demon particle distribution, $P(E_d,N_d)$,
for $E_d = 0$ ($\circ$), 3 ($\square$), and 6 ($\diamond$) with
corresponding slopes $-2.85$, $-2.83$, and $-2.76$, respectively.
The differences in the various slopes is a measure of the error due
to the limited duration of the simulations and finite size effects.} 
\end{center}
\end{figure}

The results for $\ln{P(E_d,N_d)}$ versus $E_d$ and $N_d$ are shown in
Fig.~\ref{fig4} for $E = 400$, $N = 200$, and 10000\,mcs. We include all
the data so the reader will know what to expect. The results for
large values of $E_d$ or $N_d$ are not accurate because they correspond
to probabilities that are too small for the simulation to
determine accurately. If we were to run significantly longer, we would
obtain more accurate values. A linear fit to those parts of the data that
fall on a straight line leads to $\beta \approx 0.26$ and $ \beta\mu \approx
-2.8$.

\begin{figure}[h]
\begin{center}
\includegraphics[scale=0.5]{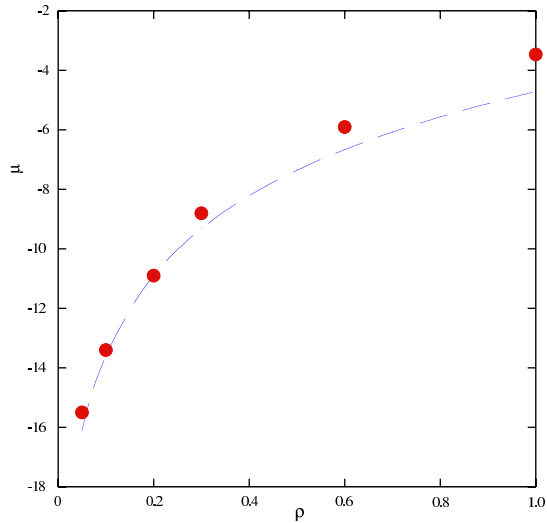}
\caption{Computed chemical potential $\mu$ of the one-dimensional ideal gas
versus density using the generalized demon algorithm. The curve is the
analytical result, Eq.~(\ref{analytic2}), for
$\mu$ assuming an ideal semiclassical gas with $\hbar = 1$.} \label{fig5} 
\end{center}
\end{figure}

We can compare these simulation results with the analytic result for a
one-dimensional ideal gas in the semiclassical limit:
\begin{equation}
\mu = -kT \ln{\Big[
\frac{L}{N}\Big(\frac{2\pi mkT}{h^2}\Big)^{1/2}\Big]}, \label{analytic}
\end{equation}
where $h$ is Planck's constant. Because the position and
momentum values are integers in our simulation and the analytical result
is derived assuming $\Delta x \Delta p = h$, we have
$h = 1$ in our units. Also, we have $m = 1/2$ and $k = 1$. Thus,
in our units Eq.~(\ref{analytic}) reduces to
\begin{equation}
\mu = -T \ln{\big[ \frac{L}{N}\bigl(\pi T
\bigl)^{1/2}\big]}. \label{analytic2}
\end{equation}
Figure~\ref{fig5} shows the chemical potential extracted from the
slopes of $\ln{P(E_d,N_d)}$ as a function of the density
$\rho = L/N$ for
$E/N = 2$. All data were generated with at least 1000\,mcs, which is
sufficient to yield $\mu$ to within a few
percent. As can be seen in Fig.~\ref{fig5}, the numerical data is
indistinguishable from the analytical results for small $\rho$, but
deviations occur for larger $\rho$, where the analytical value for the
chemical potential in Eq.~(\ref{analytic2}) is less than the value from the
simulations. The reason is that although the analytical
calculation includes the same states as the lattice model, the analytical
result also includes
states with multiple occupancy, which become important as the density
increases. Because the lattice model has fewer states than is included in
the analytical calculation, it is more difficult to add a particle to the
system, and we should obtain a larger chemical potential (less
negative). The increased difficulty in adding particles to the system
leads to a higher probability that the demon will have more particles
and thus a shallower slope for
$\ln{P(E_d,N_d)}$ versus $N_d$.

We can test this explanation further by changing our simulation so that
more than one particle is allowed at any site in phase space. Because the
analytical calculation undercounts such states, we expect our
simulation to lead to a lower value for the chemical potential than the
analytical value. For a dense system with $L = 1000$,
$N =1000$, and $E= 2000$, we find $T = 4.1$ and $\mu = -6.2$, which is
lower than the analytical value of
$-5.2$. However, for the same conditions with only single particle
occupancy allowed, we find $T = 3.7$ and $\mu = -3.5$, which is larger
than the analytical value of
$\mu=-4.5$ for this value of $T$.

A negative chemical potential means a negative slope for
$\ln{P(E_d,N_d)}$ versus $N_d$. Again we can understand why the chemical
potential must be negative. If it were not, then the probability that
the demon has
$N_d$ particles would increase with increasing $N_d$ instead of
decreasing. Although particles can easily move into the demon because it
lowers the system's energy to do so, there are many places in phase
space for the demon to return particles to the system if the density is
small. Any energy the demon receives from taking in a particle can be
used to return the particle to the system. Thus, the demon and the
system will quickly come to equilibrium such that the mean number of
demon particles is much less than the mean number of particles in the
system. However, if the system is very dense, then the system acts as a
dense Fermi gas at low temperatures for which the chemical potential
should be positive.

\begin{figure}[h]
\begin{center}
\includegraphics[scale=0.5]{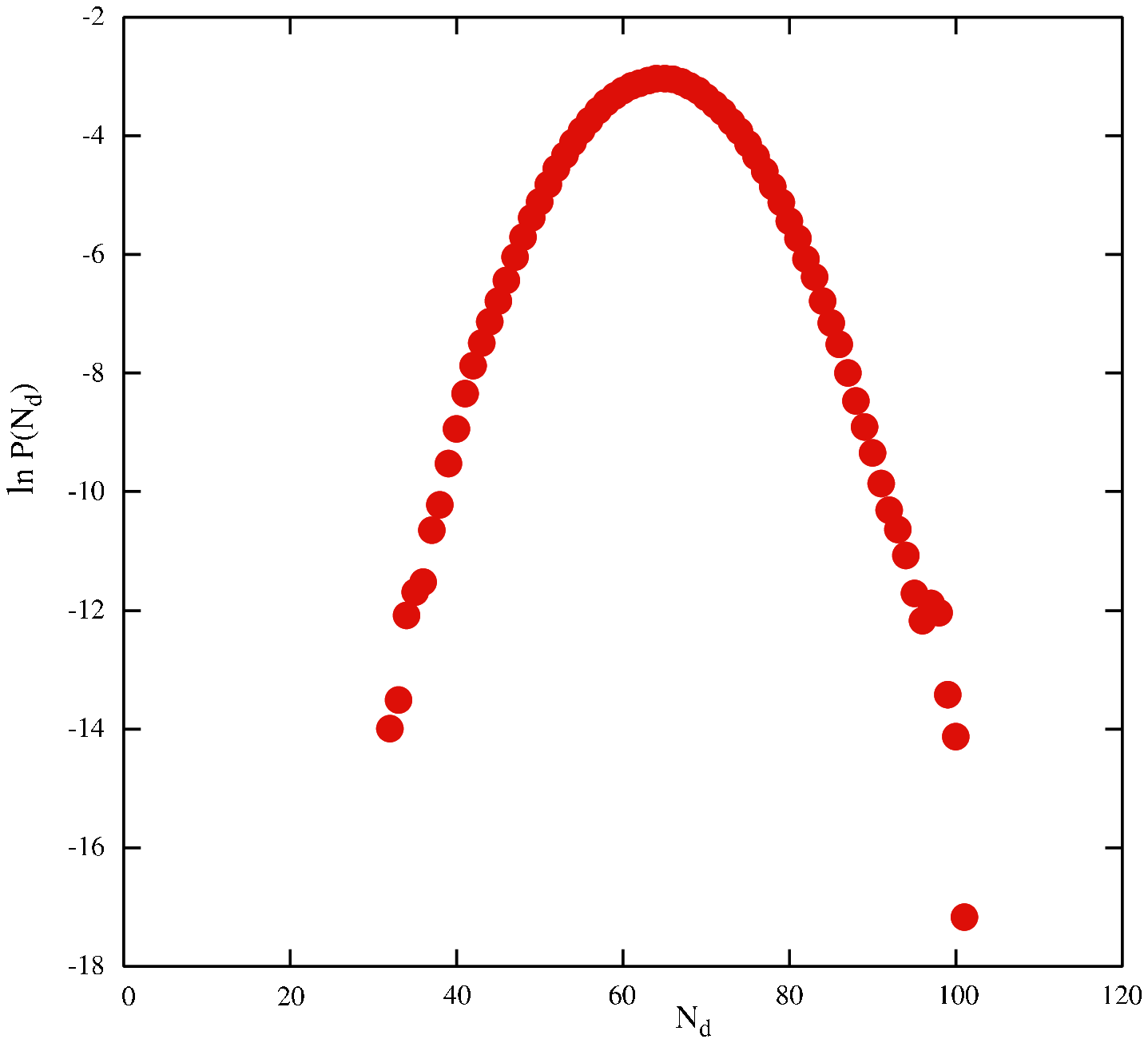}
\caption{The demon particle distribution, $P(E_d=0,N_d)$, for a dense
one-dimensional ideal gas system with $N = 200$, $E = 50$, and $L =
200$. Because the chemical potential is
positive in this case, the demon takes on a significant fraction of the
particles.} \label{fig6} 
\end{center}
\end{figure}

At high densities and low energies, it becomes very difficult to find a
place in phase space to return a particle to the system, and hence the mean
number of particles in the demon is a significant fraction of all the
particles, and is no longer a small perturbation on the system. Thus we
would not expect to find an exponential distribution, and we instead find a
Gaussian distribution as shown in Fig.~\ref{fig6}. (A parabolic plot for
$\ln P$ indicates a Gaussian distribution for 
$P$.) However, even though the demon takes on many particles, most of them
have zero energy, and the plot of $\ln P$ versus $E_d$ is still linear. For
this particular simulation the temperature was
$T = 0.49$. If the chemical potential were negative, then near $N_d = 0$ the
slope of $\ln P{(N_d)}$ would be negative. In our simulation it is positive
which indicates a positive chemical potential. 

Our simple ideal gas lattice model also provides another example of how
the equipartition theorem is not always applicable. For a relatively long
run of 32,000\,mcs with $E = 200$ and
$N=100$, the average energy per particle in the system was found
to be 1.965 and the temperature measured by the demon was 3.76. (The demon
holds part of the total energy, so the average energy per
particle in the system must be less than
$200/100 = 2$.) By the equipartition function, each degree of freedom
provides
$(1/2)T$ in energy, and thus the temperature is given by
$\half T = 1.965$ or $T= 3.93$. The difference between this predicted
value and the simulation value is statistically significant and is due to
the discrete nature of the possible energy values in the simulation. 

\subsection{Effects of interactions} 

We next discuss the effects of including interactions between the
particles in the system and consider a lattice-based model with a hard
core repulsion so that no two particles can be at the same position even
if they have different momenta. We also consider a model with a hard core
and an attractive square well interaction between nearest neighbor
particles with an energy depth equal to
$-1$. We will call these models the hard core and
square well models, respectively. Our results are shown in
Table~\ref{table1}. For all these runs at least 1000\,mcs were done.

For dilute systems,
$N = 100$ or 
$\rho = N/L = 0.1$, we see that the temperature is the same, $T \approx
3.8$, for all three systems with the same initial energy per particle, $E/N
= 2$. For the denser system with $\rho = 0.6$, the three models have
different temperatures because the interactions are now important. The
difference between the ideal and hard core gas is small. However, the
temperature is significantly higher for the square well model. The reason
is that if there is an attractive square well, some of the total energy is
(negative) potential energy, and thus the total kinetic energy must be
greater to maintain the same sum of kinetic and potential energies as
for the other two systems.

The chemical potentials shown in Table~\ref{table1} are approximately
the same for the dilute low temperature ideal gas and hard core systems,
but the existence of the attractive well lowers the chemical potential
slightly. The reason is that adding a particle can lower the energy if
there is an attractive square well. As
the temperature increases, this effect becomes less important because
the kinetic energy dominates the potential energy.

In the dense system,
$\rho = 0.6$, the differences between the three models become more
pronounced. Adding a hard core repulsion increases the chemical
potential, because the hard core makes it more
difficult for the system to accept particles compared to the ideal gas.
When we add the attractive square well, the energy for
particles to be added next to existing particles is lowered, thus making
it easier for the system to gain particles. The result is a lowering of
the chemical potential compared to the hard core model, which is what we
find in our simulations. Thus, our results confirm the idea that the
chemical potential measures how easy it is for a system to accept
particles.

\subsection{Off-lattice simulations}

We can do analogous simulations without a lattice using a similar
procedure. Instead of visiting a lattice site and determining if there
is a particle present, we introduce a small volume in phase space,
$V_p$, choose a location in phase space at random, and determine if there
is a particle in this volume centered at the chosen location. The
results for the chemical potential will depend on the value of $V_p$, in
analogy to choosing a value for the cell size in the analytical
calculations. However, this dependence is not important because only
differences in the chemical potential are physically meaningful.

To compare with our previous results, we simulated a one-dimensional hard
core gas including the momentum degrees of freedom. We set
$V_p = 1$ to be consistent with the value used in our lattice
simulations. Because we used a spherical volume (a disk in
two-dimensional phase space), the radius of the disk was chosen to be 
$1/\sqrt{\pi}$, so that its area equals unity. For $N = 100$ and $E =
200$ we obtained the same results as with a lattice system within
statistical errors.

A common inter-particle potential used to model the
interaction of simple atoms is the Lennard-Jones potential defined
by
\begin{equation} V(r)= -4
\epsilon\big[(\frac{\sigma}{r})^{12} - (\frac{\sigma}{r})^6 \big],
\label{lj}
\end{equation}
where $\sigma$ (equal to $3.4 \times 10^{-10}$\,m for argon) is the diameter of the repulsive part of the
potential and $\epsilon$ (equal to $1.65 \times 10^{-21}$\,J for argon) is
the depth of the attractive part. The potential is cutoff at $3\sigma$ to
reduce the number of interactions. We choose units such that $\sigma = 1$ and
$\epsilon = 1$. We did a simulation at low temperature with parameters
$N = 200$, $L = 1000$, and a momentum axis ranging from $-1$ to $+1$. This
range for the momentum axis is sufficiently large because we initialized
the system by randomly placing particles in phase space so that the total
energy did not exceed $E = 1$. After about 2000\,mcs we found the system
energy to be $-0.39$, corresponding to a temperature of 0.60 (corresponding to 72\,K for argon) from the mean
demon energy and a chemical potential of
$-0.29$ (corresponding to  $-4.8 \times 10^{-22}$\,J or about 0.003\,eV) from the demon particle distribution. The chemical potential is close
to zero because the total system energy is so low that the amount of phase
space available for adding a particle is small. This explanation for the value of
the chemical potential reminds us that the chemical potential is not just a
measure of the energy needed to add a particle, but also is a measure of the
number of available states for adding a particle.

\section{Conclusion}
The usual demon algorithm and its generalization
can be used to help students understand the concepts of
temperature and chemical potential. We believe that discussions of these
algorithms in a thermal and statistical physics class can give students a
more concrete model of thermal and diffusive interactions than is possible
by formal mathematical derivations. Although students will gain even more
understanding by writing and running their own programs, much of the benefit
can be obtained by discussing the algorithms and how they lead to a measure
of the temperature and chemical potential. It also is important to
ask students to describe how the demon's energy and particle number
distributions will change under different circumstances.

\begin{acknowledgements}
Gould and Tobochhnik acknowledge support from the National
Science Foundation under award PHY-98-01878, and Machta acknowledges
support from DMR-0242402. Tobochnik also acknowledges useful discussions
with Michael Creutz. 
\end{acknowledgements}

\section{Appendix: Suggestions for further study}

In the following we suggest a few problems for students. Example programs
and their source code (in Java) can be downloaded from {\tt
{<}stp.clarku.edu/simulations{>}}.

\medskip\noindent Problem 1. Write a program to estimate the chemical
potential of an ideal classical gas in one dimension by using the
Widom insertion method, Eq.~(\ref{insertion}), and a
two-dimensional phase space lattice. Use the parameters $L =
1000$, and $p_{\max} = 10$, and
explore the dependence of the chemical potential on the temperature for
a fixed density. Try
$\rho = 0.2$ ($N = 200$) and
$\rho = 0.8$ ($N = 800$).

\medskip\noindent Problem 2. Write a program to simulate the
one-dimensional ideal gas using a two-dimensional phase space lattice.
Explore the dependence of the chemical potential on the temperature for
a fixed density. Try $\rho = 0.1$ and $\rho = 0.8$. Compare your results
with the analytical predictions and discuss the behavior of the
chemical potential behaves as a function of temperature and density.

\medskip\noindent Problem 3. Write a program to simulate the
one-dimensional Lennard-Jones fluid using a two-dimensional phase
space. Compute the chemical potential as a function of temperature and
density and compare your results with the corresponding results for the
ideal gas.

\medskip\noindent Problem 4. Write a program to simulate the
two-dimensional ideal gas using a four-dimensional phase space lattice.
Choose $L = 20$, but otherwise use similar parameters as those for the
previous problems. Compare your
results with the analytical ones. How does Eq.~(\ref{analytic2}) change
for two dimensions? Are there any qualitative differences between one
and two dimensions? What happens if interactions are added? Add a hard
core and attractive well. To speed up the simulation and keep the
temperature low, you can reduce the size of the momentum axes.

\newpage

\begin{table}[h]
\begin{center}
\begin{tabular}{|r|r|r|r|l|}
\hline
$N$ & $E$ & $T$ & $\mu$ & model \\
\hline
\vline 100	& 200	& 3.81 &	$-13.4$	&hard core\\
\hline
100	& 200	& 3.83	& $-13.4$	& ideal gas\\
\hline
100	& 200	& 3.91	& $-14.9$	& square well\\
\hline
100	& 800	& 15.5	& $-66.3$	& ideal gas\\
\hline
100	& 800 &	16.0	& $-66.9$	& hard core\\
\hline
100	& 800	& 15.9	& $-69.7$	& square well\\
\hline
600	& 1200 &	4.03	& $-3.45$	& hard core\\
\hline
600	& 1200	& 3.74	& $-5.90$	& ideal gas\\
\hline
600	& 1200 &	5.26	& $-6.41$	& square well\\
\hline
600	& 1000 &	4.65	& $-5.52$	& square well\\
\hline
600	& 800 & 4.01	& $-4.75$	& square well\\
\hline
\end{tabular}
\caption{\label{table1}Comparison of different one-dimensional
systems simulated on a phase space lattice. The system types are
an ideal gas with no interparticle interactions; a hard core interaction
such that no two particles can be in the same spatial position; and
a hard core plus a nearest neighbor
attractive square well with unit energy depth. $N$ and $E$ are the initial
total number of particles and energy, respectively; $T$ and
$\mu$ are computed from the slopes of $\ln P(E_d,N_d)$. In all
the simulations $L = 1000$.}
\end{center}
\end{table}

\newpage

\section*{Figure Captions}

\begin{thebibliography}{33}
\bibitem{comment}If the demon energy is not continuous, then the
initial demon energy should be consistent with the possible
energies of the demon.
\bibitem{footexplain}The mean demon energy is given
by
$\lb E_d
\rb =
\!\int_0^\infty\! E_d\,e^{-\beta E_d}/\!\int_0^\infty e^{-\beta
E_d}=kT$ if the possible demon energies are continuous. The upper
bound can be taken to be infinite even though the total energy in
our simulations is finite because the high energy states contribute
a negligible amount to the integral.
\bibitem{foot2}In this case the possible demon energies are not continuous
and $\lb E_d \rb \neq kT$ except in the limit of high temperature.
\end{thebibliography}
\end{document}